# Optical microscopy-based thickness estimation in thin GaSe flakes

Wenliang Zhang[1], Qinghua Zhao[1,2,3], Sergio Puebla[1], Tao Wang[2,3,*], Riccardo Frisenda[1,*], Andres Castellanos-Gomez[1,*]

[1]Materials Science Factory. Instituto de Ciencia de Materiales de Madrid (ICMM-CSIC), Madrid, E-28049, Spain.

[2]State Key Laboratory of Solidification Processing, Northwestern Polytechnical University, Xi'an, 710072, P. R. China

[3]Key Laboratory of Radiation Detection Materials and Devices, Ministry of Industry and Information Technology, Xi'an, 710072, P. R. China

taowang@nwpu.edu.cn; riccardo.frisenda@csic.es; andres.castellanos@csic.es

ABSTRACT

We have implemented three different optical methods to quantitatively assess the thickness of thin GaSe flakes transferred on both transparent substrates, like Gel-Film, or $SiO_2$/Si substrates. We show how their apparent color can be an efficient way to make a quick rough estimation of the thickness of the flakes. This method is more effective for $SiO_2$/Si substrates as the thickness dependent color change is more pronounced on these substrates than on transparent substrates. On the other hand, for transparent substrates, the transmittance of the flakes in the blue region of the visible spectrum can be used to estimate the thickness. We find that the transmittance of flakes in the blue part of the spectrum decreases at a rate of 1.2%/nm. On $SiO_2$/Si, the thickness of the flakes can be accurately determined by fitting optical contrast spectra to a Fresnel law-based model. Finally, we also show how the quantitative analysis of transmission mode optical microscopy images can be a powerful method to quickly probe the environmental degradation of GaSe flakes exposed to aging conditions.

Mechanical exfoliation has proven to be a powerful technique to produce high quality 2D materials.[1–4] Unfortunately, the mechanical cleavage of a bulk layered material typically yields randomly distributed flakes with different thicknesses. Therefore, fast, reliable, and non-destructive methods to identify 2D materials and to allow distinguishing ultrathin flakes from bulky ones are crucial for the development of this line of research. This is particularly true for the case of 2D semiconductors materials where the reduction in thickness yields quantum confinement effects that have a strong impact in their band structure and concomitantly on their band gap.[5–9] Before the introduction of mechanical exfoliation in 2004 by Novoselov and Geim,[1] atomic force microscopy (AFM) and scanning electron microscopy (SEM) were the most extended methods to characterize the morphology of micro-structures and nanomaterials. Their application to identify and select 2D materials, however, resulted problematic. Mechanical exfoliation may produce a handful of suitable flakes per $cm^2$ and AFM scanning speed prevents its use from being optimal for the blind search of 2D materials. While SEM is a fast technique with a large throughput, its use tends to form a layer of contaminants on the surface of the sample.[10] This is critical for 2D materials that are in essence 'all-surface' and thus are strongly modified by the presence of adsorbates.[10] Moreover, SEM contrast does not provide a trivial way to distinguish bulk-like from ultrathin flakes.[11,12]

Shortly after the isolation of graphene by mechanical exfoliation, optical microscopy was proposed as an alternative technique to identify and estimate the number of layers of mechanically exfoliated flakes.[13–16] Since then, the isolation of a new 2D material is always followed by experimental efforts to develop optical microscopy-based recipes to identify and determine the number of layers of that novel 2D material.[17–25]

In this work, we provide a set of simple optical microscopy methods to assess the thickness of GaSe flakes. GaSe is a material of the layered metal-monochalcogenide III–VI semiconductor family and it holds the great possibilities in optoelectronics and nonlinear optics because of the fast photoresponsivity, high carrier mobility and nonlinear optical properties.[26–34] Our approach is based on the combination of quantitative analysis of the apparent color of different GaSe flakes, of transmission mode optical microscopy images and micro-reflectance spectra. We found that the apparent color of GaSe flakes, both deposited on transparent substrates or onto $SiO_2/Si$ substrates, can be used as a coarse method to assess the thickness of flakes at glance. The quantitative analysis of transmission mode optical images provides a more quantitative way to determine the thickness of GaSe flakes as the blue channel transmittance drops by ~1%/nm. Interestingly, we observed how this analysis also provides a tool to probe the degradation of this 2D material. We additionally show how a Fresnel-law based model can be used to fit the optical contrast of GaSe flakes and to determine the thickness in a more accurate way. We believe that this work can be useful to speed up future works on this emerging 2D material.

GaSe flakes were mechanically exfoliated with Nitto tape (SPV 224) from bulk GaSe ingots grown by the Bridgman technique. A thorough characterization of the as-grown bulk crystals can be found in Ref. [35]. GaSe crystallites on tape are then transferred onto the surface of a Gel-Film (Gel-Pak WF x4 6.0 mil) substrate.

The surface of the Gel-Film substrate, containing the transferred GaSe flakes, is inspected under an optical microscope (Motic BA 310-Met-T) in transmission mode to identify the GaSe flakes. **Figure 1a** shows a collection of transmission mode optical microscopy images of GaSe flakes of different thicknesses (as inferred from the different opacity) on the surface of a Gel-Film substrate. The GaSe flakes can be transferred from the Gel-Film substrate onto another target substrate by an all-dry deterministic placement method.[36–38] **Figure 1b** shows optical microscopy images of the GaSe flakes shown in (a) after being transferred onto a Si substrate with 270 nm of $SiO_2$ capping layer. The topography of these flakes can be studied by AFM (**Figure 1c**), allowing to correlate the apparent colors in (a) and (b) with their actual thickness. The color-charts included in **Figure 1a and 1b** can be therefore used to coarsely estimate the thickness of a certain GaSe flake at first glance. Note that the thickness dependent color change is sizeably weaker onto the Gel-Film substrate and thus this method can only yield a rough thickness estimation. On $SiO_2$/Si substrates, on the other hand, one can estimate the thickness of a GaSe flake with ± 3 nm accuracy.

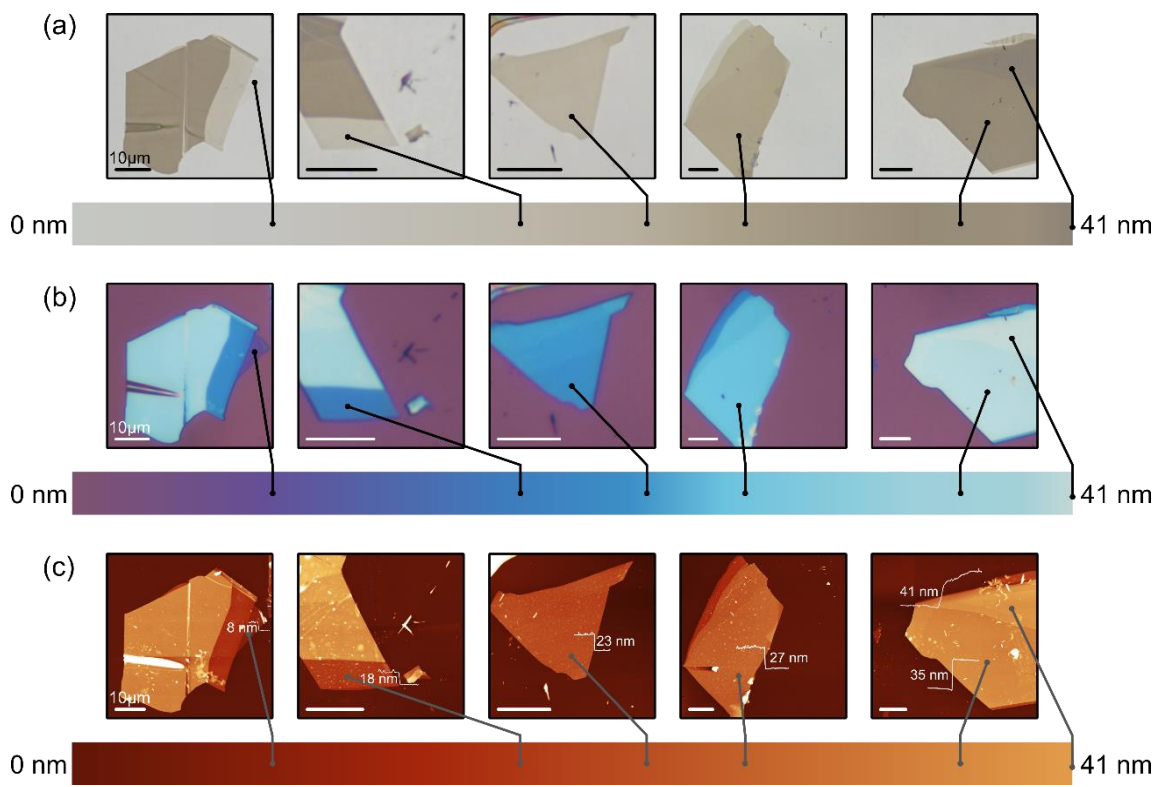

**Figure 1. Thickness dependent apparent colors of thin GaSe flakes.** (a) Transmission mode optical microscopy image of thin GaSe flakes with different thicknesses deposited onto a Gel-Film substrate. (b) Epi-illumination (reflection mode) optical microscopy images of the same GaSe flakes after being transferred onto a 270 nm $SiO_2$/Si substrate. (c) Atomic force microscopy (AFM) measurements of the corresponding GaSe flakes with thickness between 8 nm and 41 nm. All the scale bars are 10 μm.

**Figure 2a and 2b** compare a transmission mode optical microscopy image of a thin GaSe flake on Gel-Film with its AFM topography after transfer the flake onto a $SiO_2$/Si surface. Transmission images, like those shown in **Figure 1a and 2a**, have been acquired with a digital CMOS camera (AM Scope 1803 MU) and thus they are composed of three channels: red (R), green (G), and blue (B). By analysing the intensity of the individual R, G, and B channel images we can then quantitatively extract the transmittance in the red, green and blue part of the visible spectrum by normalizing the image intensity to the intensity of the bare, uncovered, Gel-Film substrate.[39] **Figure 2c** compares the transmittance extracted from the R, G, and B channels along the arrow in **Figure 2a** and the AFM topography line profile measured along the arrow in **Figure 2b**. The clear correlation between the flake transmittance and its thickness, measured by AFM, indicates that one can use the transmittance calculated from the R, G, and B channels to determine the thickness of GaSe flakes. The AFM height profile shows some 'spikes' which we attribute to the beginning of the GaSe environmental degradation process which typically occurs in the first 1-5 hours.[40]

**Figure 2d** displays the relationship between the transmittance of the B channel and the thickness of GaSe flakes. We have measured 23 flakes with thickness ranging 8−39 nm. Interestingly, within the studied thickness range the transmittance of the B channel follows a linear trend with the flake thickness with a slope of −1.2% per nm. Note that the difference between the tip-flake and tip-substrate interactions can lead to large thickness discrepancy when testing very thin flakes and introduce artifacts in the measurement.[8,41] This could explain that the observed linear trend in Figure 2d does not cross the vertical axis though 1 (100% transmission) in the limit of zero thickness.

**Figure 2e** compares the thickness values determined by AFM measurement and the values calculated through the optical transmission of B channel images. The resulting plot presents a fit line with a slope of $1 \pm 0.03$ (a perfect agreement should yield a slope of 1), illustrating the feasibility to assess the thickness of GaSe flakes on transparent substrates by quantitatively analysing transmission mode optical microscopy images.

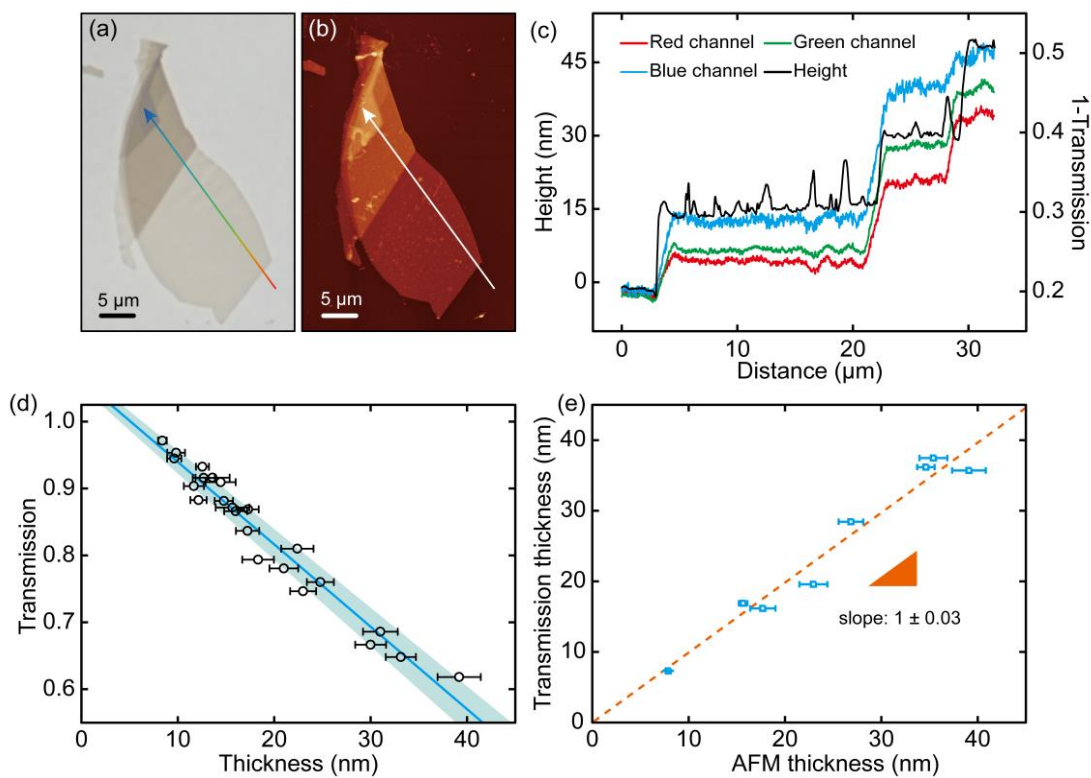

**Figure 2. Analysis of transmission images of thin GaSe flakes.** (a) Optical microscopy image of an exfoliated thin GaSe flake on a Gel-Film substrate. (b) AFM image of the GaSe flake after being transferred to a $SiO_2/Si$ substrate. The two profile lines (colored and white) in (a) and (b) correspond to the transmission and height profiles shown in panel (c). (d) Relationship between the GaSe flake thickness (determined by AFM measurement) and the optical transmission (blue channel) showing a linear trend with slope of $-(1.2 \pm 0.06)$ %/nm. (e) Comparison between the thickness determined by AFM measurement and the thickness determined through the transmission of the blue channel. The experimental data is represented using the blue squares with the error bar showing the uncertainty of the AFM measurement, the resulting fit line (orange) with a slope of $1 \pm 0.03$ demonstrates the reliability of the transmission-based thickness determination.

The analysis of optical contrast of flakes deposited on $SiO_2/Si$ substrates through a Fresnel law-based model [4,13] has been proven as an efficient way to assess the thickness of other 2D materials such as graphene [16,42,43], antimonene [44], transition metal dichalcogenides [17,18,45,46], mica [19], and hexagonal boron nitride [47]. In the following we will adapt this method to study GaSe flakes. **Figure 3a** shows a schematic drawing of a four-media (air/GaSe/SiO$_2$/Si) optical system. The optical contrast ($C$) of the GaSe flake is defined as:[13]

$$C = \frac{R_{\text{GaSe}} - R_{\text{sub}}}{R_{\text{GaSe}} + R_{\text{sub}}} \tag{1}$$

being $R_{\text{GaSe}}$ and $R_{\text{sub}}$ the light intensity reflected by the GaSe flake and the substrate respectively. We acquired reflectance spectra under normal incidence with a modified metallurgical microscope (BA 310 MET-T, Motic) whose experimental operation details have been described in our previous works.[48,49]

The optical contrast of this kind of multilayer system can be calculated by accounting for the light reflected and transmitted at each different interface with a Fresnel law based model.[4,13] The reflection coefficient in a Fresnel model with four media can be expressed as:[4,13]

$$r_{\text{GaSe}} = \frac{r_{01}e^{i(\Phi_1+\Phi_2)}+r_{12}e^{-i(\Phi_1-\Phi_2)}+r_{23}e^{-i(\Phi_1+\Phi_2)}+r_{01}r_{12}r_{23}e^{i(\Phi_1-\Phi_2)}}{e^{i(\Phi_1+\Phi_2)}+r_{01}r_{12}e^{-i(\Phi_1-\Phi_2)}+r_{01}r_{23}e^{-i(\Phi_1+\Phi_2)}+r_{12}r_{23}e^{i(\Phi_1-\Phi_2)}}, \tag{2}$$

where the subscript 0 refers to air, 1 to GaSe, 2 to SiO$_2$ and 3 to Si. Under normal incident light, the phase shift induced by the propagation of the light beam in the media $i$ is $\Phi_i = 2\pi\tilde{n}_i d_i/\lambda$, with $\tilde{n}_i$, $d_i$ and $\lambda$ representing the complex refractive index, thickness of the medium $i$ and the wavelength, respectively. The coefficient $r_{ij} = (\tilde{n}_i - \tilde{n}_j)/(\tilde{n}_i + \tilde{n}_j)$ is the Fresnel coefficient at the interface between the media $i$ and $j$.

The reflection Fresnel coefficient in a three media, the case of the bare substrate without being covered by GaSe flake is expressed as:

$$r_{\text{sub}} = \frac{r_{01} + r_{12}e^{-i2\Phi_1}}{1 + r_{01}r_{12}e^{-i2\Phi_1}} \tag{3}$$

Where sub index 0 is air, 1 is SiO$_2$ and 2 is Si. Using equations (2) and (3), we can calculate the optical contrast by firstly calculating the reflected intensity of both situations as

$$R_{\text{GaSe}} = |\overline{r_{\text{GaSe}}} r_{\text{GaSe}}|, R_{\text{sub}} = |\overline{r_{\text{sub}}} r_{\text{sub}}|. \qquad (4)$$

Then the optical contrast can be obtained through the following equation (5) that correlates the reflected intensity by the bare substrate ($R_{\text{sub}}$) with the reflected intensity by the GaSe flake ($R_{\text{GaSe}}$) as:

$$C = \frac{R_{\text{GaSe}} - R_{\text{sub}}}{R_{\text{GaSe}} + R_{\text{sub}}}. \qquad (5)$$

We found that plugging into equations (1) to (5) the thickness of the SiO$_2$ layer (determined by reflectometry), the thickness of the GaSe flakes (measured with AFM), the reported refractive indexes for air, SiO$_2$ and Si, and assuming a thickness-independent refractive index $\tilde{n}_i = 2.5 - i0$ for the GaSe flakes, compatible with the literature values that are in the ~2.1-2.9 range,[50,51] we can reproduce the experimental optical contrast spectra accurately (see **Figure 3b**). Moreover, one can determine the thickness of a GaSe flake directly from a measured optical contrast spectrum by calculating optical contrast spectra for different GaSe thicknesses and computing which one provides the best matching with the experimental data. **Figure 3c** compares the experimentally measured optical contrast of a GaSe flake (24 nm thick according to the AFM measurement) with the modelled optical contrast assuming a thickness in the range of 20−28 nm. There is a clear best match for a thickness of 23.5 nm, in good agreement with the AFM value, illustrating the feasibility to assess the thickness of GaSe flakes with this method. The inset in **Figure 3c** shows the square of the difference between the measured contrast and the calculated one as a function of the thickness. The plot shows a well-defined minimum centered at a thickness of 23.5 nm. We estimate that this method provides an uncertainty in the thickness determination in the 1−2 nm range. In **Figure 3d** we compare the flake height values determined with AFM with the values obtained following the discussed optical contrast fit method. In this plot, one can find a slope value of 0.95 ± 0.05 marked by the straight line, which indicates the good agreement between the thicknesses of thin GaSe flakes measured by AFM and the fit to the Fresnel law-based model.

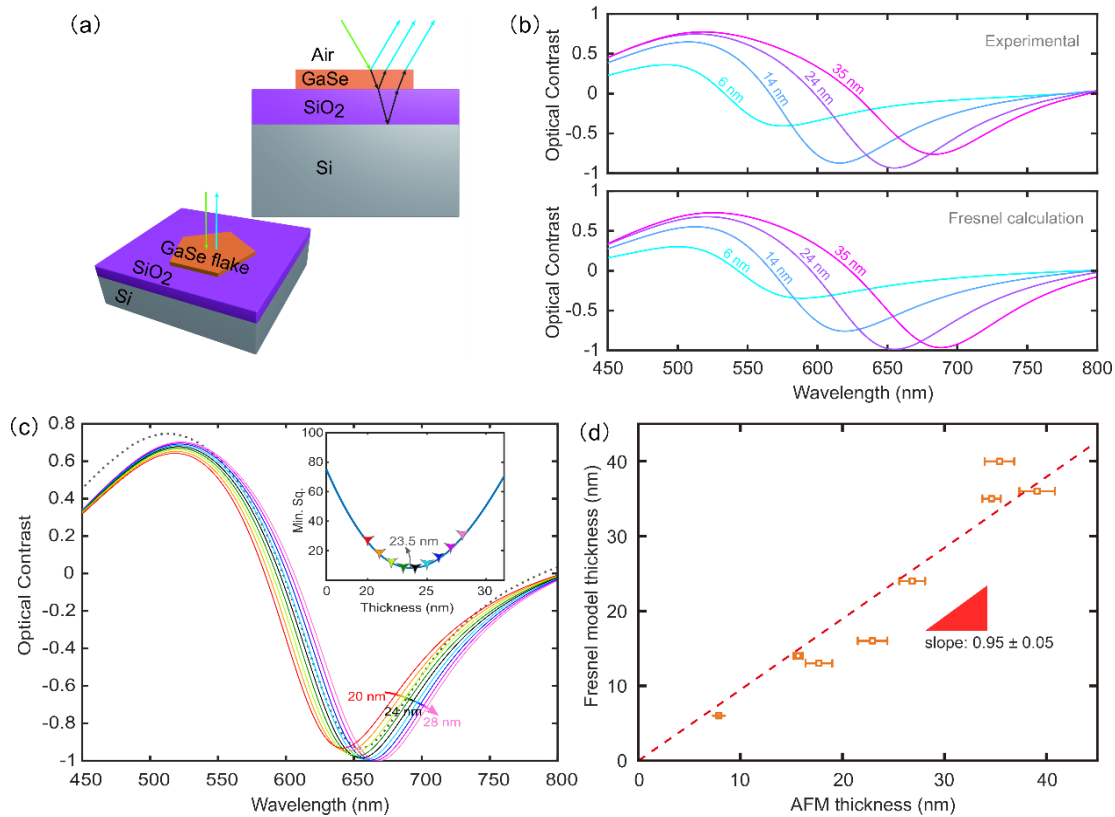

**Figure 3. Optical contrast analysis of thin GaSe flakes on SiO₂/Si substrates.** (a) Fresnel model based on a four-media system (air/GaSe/SiO$_2$/Si) utilized for calculating the optical contrast. (b) Experimental optical contrast spectra (top panel) and the calculated ones based on Fresnel model (bottom panel) of thin GaSe flakes with different thickness on 270 nm SiO$_2$/Si substrates. (c) Comparison of experimental optical contrast spectrum (black dot line) of a ~24 nm thick GaSe flake deposited on a 270 nm SiO$_2$/Si substrate and the calculated ones (colored solid lines) assuming a flake thickness from 20 nm to 28 nm. The inset shows the minimum square value at different flake thickness. (d) Comparison between the thickness determined by AFM measurement and the Fresnel model. The experimental data is represented using the orange squares with the error bar showing the uncertainty of the AFM measurement, the resulting fit line

(red) with a slope of 0.95 ± 0.05 demonstrates the reliability of the thickness assessment by fitting the optical contrast to the Fresnel law based model.

Several previous works have reported how GaSe tend to degrade upon air exposure. [40,52–57] These works used atomic force microscopy, Raman spectroscopy and electrical transport measurements to probe the flake degradation process. It is interesting to check whether a simpler optical microscopy-based technique could be used to monitor aging effects on GaSe flakes. **Figure 4a** shows a time sequence of transmission mode optical microscopy images of a GaSe flake on a Gel-Film substrate. The flake has been exposed to air (at room temperature, 23°C, with a relative humidity ranging ~20-40%) and illuminated with light with 420 nm wavelength (with a power density of ~16 mW/mm$^2$) to accelerate the aging effects on the flake. **Figure 4b, 4c and 4d** display the time evolution of the transmittance in the R, G, and B channels showing an increase of transmittance during the first ~5 hours. This increase in transmittance could be due to a reduction of the flake thickness (i.e. the flake is being etched away during the aging) or by the transformation into other chemical compounds with lower optical adsorption. **Figure 5a and 5b** show two AFM images of a GaSe flake on SiO$_2$/Si before and after subjecting it to an aging process like that described above. The topography (**Figure 5c**) shows how the flake has not been etched away, on the contrary its thickness has increased. This observation points out that the increased transmittance upon aging must be due to the transformation of GaSe into more transparent chemical compounds, containing Ga$_2$Se$_3$, Ga$_2$O$_3$, and/or amorphous selenium (a-Se), in agreement with previous Raman measurements.[40,53–55,58] The topography of the flake has also changed substantially, becoming much rougher than the pristine flake. It seems that the degradation is accompanied by the generation of 'bubbles' or 'blisters' on the surface.

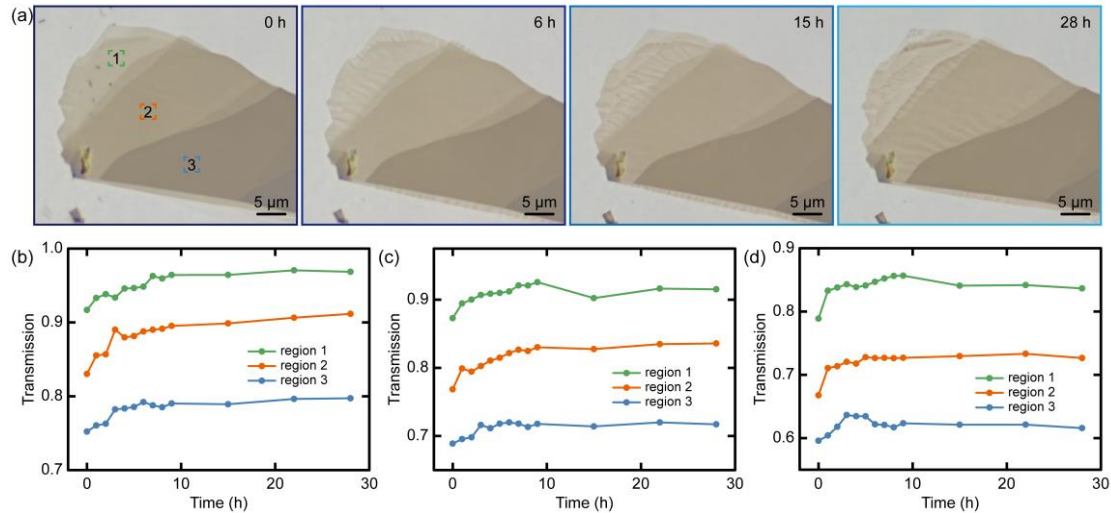

**Figure 4. Time evolution of the optical transmission of the thin GaSe flake during the photodegradation with the illumination of 420 nm light.** (a) Sequence of transmission optical microscopy images of a thin GaSe flake at different degradation time under a 420 nm UV light. (b-d) The optical transmission of different regions, corresponding to the colored squares in the images of panel (a), recorded as a function of degradation time upon air exposure for red channel (b), green channel (c), and blue channel (d), respectively.

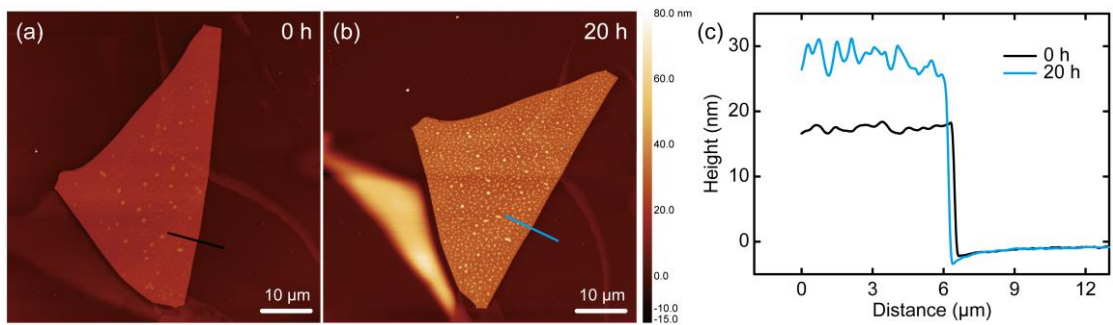

**Figure 5. The change of thickness of the thin GaSe flake upon air exposure under the illumination of 420 nm light.** (a, b) AFM images of a pristine GaSe flake (a) and the aged flake after 20 h of 420 nm illumination (b). (c) Comparison between the thickness measured on a pristine flake and the aged flake.

CONCLUSIONS

In summary, we have adapted several optical microscopy-based methods to evaluate the number of layers of thin GaSe deposited on different substrates. We demonstrate how one can quickly estimate the GaSe flake thickness with an error of ± 3 nm directly from the observed apparent color of the flake based on a calibrated color versus thickness chart. For flakes deposited on transparent substrates (like Gel-Film), we show how the transmittance in the blue part of the visible spectrum decreases at a rate of 1.2 %/nm. We also show how fitting experimentally acquired optical contrast spectra to a Fresnel law-based model one can determine the thickness of GaSe flakes that have been deposited on $SiO_2$/Si substrates. Finally, we explored the use of optical microscopy to probe the environmental degradation of GaSe finding that a quantitative analysis of transmission mode optical microscopy pictures can be used to infer the degradation of GaSe flakes.

**ACKNOWLEDGEMENTS**

This project has received funding from the European Research Council (ERC) under the European Union's Horizon 2020 research and innovation program (grant agreement n° 755655, ERC-StG 2017 project 2D-TOPSENSE) and the European Union's Horizon 2020 research and innovation program under the Graphene Flagship (grant agreement number 785219, GrapheneCore2 project and grant agreement number 881603, GrapheneCore3 project). This work received funding from the Spanish Ministry of Economy, Industry and Competitiveness (MINECO) through the grant MAT2017-87134-C2-2-R. R.F. acknowledges the support from the Spanish Ministry of Economy, Industry and Competitiveness (MINECO) through a Juan de la Cierva-formación fellowship 2017

FJCI-2017-32919. W. Zhang acknowledges the grant from China Scholarship Council (CSC) under No. 201908610178. Q.H.Z. acknowledges the grant from China Scholarship Council (CSC) under No. 201706290035.

**DATA AVAILABILITY:**

The raw/processed data required to reproduce these findings cannot be shared at this time due to technical or time limitations.

Data will be made available on request.